\documentclass[reprint, aps, amsmath, amssymb, superscriptaddress, floatfix]{revtex4-2}

\usepackage{graphicx}
\usepackage{dcolumn}
\usepackage{bm}
\usepackage{color}

\begin{document}

\title{Electron-detachment cross sections for O$^{-}$ + N$_{2}$ near the \\ 
free-collision-model velocity threshold.}

\author{A. A. Mart\'inez}
\affiliation{Instituto de Ciencias F\'isicas. Universidad Nacional 
             Aut\'onoma de M\'exico, Apartado Postal 48-3, Cuernavaca 
             62251, Mexico.}
\author{M. M. Sant’Anna}
\affiliation{Instituto de F\'isica, Universidade Federal do Rio de Janeiro,
             Caixa Postal 68528, Rio de Janeiro, 21941-972, RJ, Brazil}
\author{G. Hinojosa}
\email{Corresponding author: hinojosa@icf.unam.mx}
\affiliation{Instituto de Ciencias F\'isicas. Universidad Nacional 
             Aut\'onoma de M\'exico, Apartado Postal 48-3, Cuernavaca 
             62251, Mexico.}


\begin{abstract}
 We present measurements of the total projectile-electron-loss cross sections
 in O$^{-}$ + N$_{2}$ collisions in the energy range from 2.5 to 8.5 keV. 
 Two different techniques, beam attenuation and the growth rate, are employed. 
 The difference between the values obtained with the two methods is explained under
 the hypothesis of a contribution from anionic metastable auto-detaching states. 
 Under this hypothesis, the long-standing question of a strong disagreement among 
 reported measurements at the low-energy range is also explained. The cross sections 
 measured using the growth-rate method show a threshold behavior. 
 We analyze the cross-section velocity dependence in the framework of a collision 
 between a quasi-free electron, loosely bound to the projectile, and the molecular 
 target. Within the free collision model, we deduce and test a simple analytical 
 expression for the expected velocity threshold taking into account the angular 
 distribution of electron velocities within the anion.
\end{abstract}

\maketitle

\section{INTRODUCTION}

 In negative ions, the extra electron is bound to its neutral core by weak 
 Coulomb-screened potentials and by electron correlation effects. For this
 reason, both structure and collision models, even for the simplest cases, 
 demand highly electronically correlated approaches and are, in turn, 
 a severe test-bed for state-of-the-art theoretical methods. 
 Such may be the case of the realization of negative methane \cite{yo2020}
 and the extremely long-lived CCH$_2^-$ anion \cite{jensen2000}. 
  
 Negative ions are, in general, an important species in Nature. Its presence in
 cold plasmas \cite{medvedev2019, mondal2019} is relevant to the
 electronic density function since they are a source of thermal electrons.
 In atmospheric environments, anions were found in the upper atmosphere of
 Titan satellite \cite{coates2007}, in the coma of comets \cite{cordiner2013},
 and are of importance in artificial atmosphere generation \cite{strogonova2019}.
 In the case of the interstellar medium (IM) \cite{fortenberry2015}, despite the
 hostile conditions for their permanence, negative ions are abundant and
 a well-established fact. Understanding their presence in the IM constitutes a 
 current question in science \cite{khamesian2016}. Even in the well-established 
 field of small-mass spectrometry, a newly negative ion species (of $m/q = -16$) 
 has been reported \cite{yo2020,hernandez2014}. These questions prompt the need 
 for more fundamental studies about this kind of ions and their interactions.
 Particularities of negative-ion projectiles, when compared to the more simple 
 positive-ion or atomic projectiles, are observed in the search for 
 scaling laws for projectile-electron-loss 
 cross sections \cite{zappa2004,santos2004,wu2007,geng2013,min2017,yang2022}.
 
Even for a projectile as simple as an electron, scattering by the N$_2$ target 
show a large variety of involved processes \cite{song2023}. 
When the projectile has internal structure, and the high degree of electronic 
correlation characteristic of anions, the complexity of the collision system 
increases significantly. In this work, we present measurements of electron-detachment  
cross sections for the O$^-$+N$_2$ collision system. The measurements support an explanation, 
based on the possible unnoticed production of autoionizing states during experiments, for a 
long-standing discrepancy among reported cross section data at energies of few keV.  

 The velocity range of the cross sections reported here allows a study of 
 the onset of a projectile-electron-loss mechanism with a quasi-free-electron 
 behavior. 
 The experimental data measured with the growth rate method present a velocity threshold. 
 We show that the threshold value is consistent with the estimate of the free collision 
 model (FCM), taking into account energy and momentum conservation and the angular dispersion
 of a quasi-free electron within the anion. A general and simple analytical expression
 for this FCM threshold velocity is presented and tested with the O$^-$+N$_2$ 
 collision system.

\section{EXPERIMENT}
 
 The experimental method used here has been described in previous publications
 \cite{hernandez2018, lira2021, vergara2021}. The method consists of the application
 of two different well-established techniques, namely, beam attenuation (BAT) and  
 signal growth rate (SGR). Both techniques are used rather than one alone. This offers 
 an improved insight, given that the physics from each technique can be different, 
 as explained ahead. In short, the method is based on measuring the remaining negative
 ions from an O$^-$ ion beam after their interaction with N$_2$ gas (BAT), or the resulting
 neutral atoms of oxygen resulting from the electron loss from the O$^-$ ion-beam 
 interaction with N$_2$ (SGR). Following, we present a self-standing description of the 
 experimental method with emphasis on the present study.

\begin{figure}
\begin{center}
 \includegraphics[width=\columnwidth]{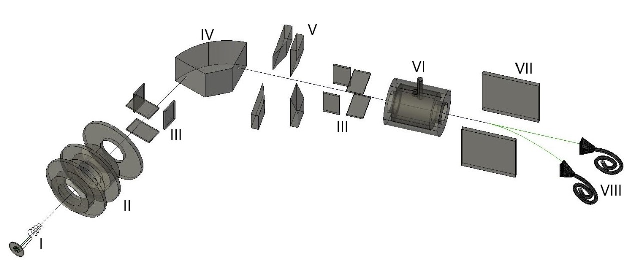}{}
  \caption{ \label{machine} Schematic diagram of the apparatus. Not presented to scale. 
  I, This filament is inside a small quartz chamber (not shown).
  II, electrostatic lens set.
  III, steering plates.
  IV, magnet sector.    
  V, collimating slits.
  VI, gas-cell.
  VII, analyzing parallel plates (PP).  
  VIII, Particle counters or channel electron multipliers (CEMs).
  }
\end{center}  
\end{figure} 

 In the initial stage, an ion beam of negative oxygen ions is produced. This was carried 
 out by introducing a combination of argon and carbon dioxide gases into a cylindrical 
 chamber made of quartz containing a tungsten filament that was heated to incandescence. 
 The pressure in this chamber was monitored not to exceed the order of 10$^{-1}$ Torr. 
 At one end of the quartz chamber, a metallic cap with a centered 1.5 mm diameter orifice 
 had a 100 V voltage applied to it in order to accelerate negatively charged particles. 
 Electrons ejected by the filament suffered this acceleration and, in turn, interacted 
 with the gas mixture producing a plasma \cite{hernandez2014} from which negative ions 
 could be extracted. The assembly containing the chamber was biased to the acceleration 
 voltage. Negative ions from the plasma were expelled through its orifice and in turn 
 repelled or accelerated toward a focusing electrostatic lens set, yielding to a magnetic 
 field where the ions are separated according to their momentum and electric charge, 
 thereby resulting in a 
 mono-energetic, mass-analyzed ion beam. After the magnetic-field region, the ion beam
 passes through a set of collimating slits to control its width and therefore its
 intensity. These slits were mounted on linear motion feed-throughs in such a way that the
 ion beam could be scanned in both directions at right angles on the normal plane to the 
 ion-beam trajectory, and hence be able to monitor its profile too.
 Before and after the magnetic field's region, two sets of parallel plates (PP), 
 installed perpendicularly to the ion-beam direction, were used to apply small electric 
 fields to further, fine-tune the ion-beam trajectory towards a gas cell. 
 The gas cell was built with input and output collimator orifices with diameters of 
 1.0 mm and 1.5 mm respectively, and with a length of 6 cm.
 
 Overall, a single species, mono-energetic, collimated, and focused beam, with a width
 of $\approx$ 0.5 mm, enters a gas cell where the interaction of O$^-$ and N$_2$ 
 occurs in a time-frame of a fraction of 10$^{-15}$ s,

\begin{equation}\label{collisEq}
 \mbox{O}^- + \mbox{N}_2 \longrightarrow \mbox{O}^* + \mbox{e}^- + \mbox{N}_2^*
\end{equation} 
 
 \noindent where the star is used to indicate an unknown final internal state. High-purity 
 N$_2$ gas was used. The resulting neutral oxygen atoms continued in their original trajectory 
 towards a channel electron multiplier (CEM) located in the symmetry axis of the apparatus.
 The remaining O$^-$ ions from the parent ion beam were separated by a perpendicular electric 
 field set to steer the ions toward a second CEM installed off-axis of the accelerator
 symmetry axis. This electric field was generated with a set of parallel plates (shown in Fig.
 \ref{machine} labeled as PP). 
 
 The time of flight of an O$^-$ ion from the output of the 
 ion source to the entrance of the gas cell was of the order of $\mu$s. The experiment
 was performed under high vacuum conditions so that the mean free path of the ion beam was,
 at all times, larger than its total trajectory.
 
 To verify if there was a full collection of charged particles, the ion-beam intensity was measured
 as a function of the PP's electric field. This produced a profile of the ion beam as
 a function of the PP voltage that systematically showed a distribution with a plateau, 
 thereby showing that the ions were fully collected because their spread was smaller than 
 the width of the CEMs acceptance aperture after dispersion in the gas-cell. The CEMs 
 collecting apertures were rectangular with dimensions of 7 mm (width) and 15 mm (height).

 As a test for the neutral atoms collection, we executed a check performed in another 
 experiment done with the same apparatus and with a lighter H$^-$ ion-beam
 \cite{vergara2021} under similar experimental conditions. In that experiment, the 
 distance between the gas cell and the detectors was changed between two experimental 
 campaigns providing a difference in the aspect ratios (detector's width to distance 
 from the gas cell) of about 18\%. This resulted in no measurable difference between 
 the two sets of electron-loss cross sections, thereby, verifying that the different 
 amounts of collected neutral atoms were not measurable.
 
 To verify that both CEMs had the same detection efficiencies, the counting rate in 
 the central CEM was measured with the PP electric field set to zero, corresponding to 
 a total count rate of residual neutral atoms plus the O$^-$ parent ion-beam counts. 
 With the PP electric field on, the lateral CEM count rate (parent ion beam) plus 
 the central CEM count rate (residual neutral atoms) was checked to add up to the 
 total count rate in the central CEM. This check was performed under empty gas-cell
 condition. The CEMs' bias voltage gains were also slightly adjusted for maximum 
 counting rates and the count rates were kept below 1$\times 10^4$ s$^{-1}$ as to 
 guarantee optimal performance of the CEMs.

 Systematic errors originate from uncertainties in gas-cell pressure and temperature
 measurements that propagate a maximum of 11\% uncertainty to the cross-section measurements. 
 The relative pressure in the gas cell was monitored with a Baratron capacitance manometer. 
 The  kinetic
 energy had a maximum of 5\% uncertainty. In the ion source chamber the base pressure 
 was 8$\times$10$^{-4}$ Pa without gas load, and 5$\times$10$^{-2}$  
 Pa with gas load. The detection chamber pressure was 5$\times$10$^{-5}$ Pa. 

 Total electron-loss cross sections were derived from the total count signal measured 
 in each of the CEMs (centered and lateral) as a function of the N$_2$ gas target thickness 
 ($\eta$). Using the signal from the neutral atoms of oxygen (centered CEM) we
 applied the signal growth rate method (SGR), and with the signal in the lateral CEM,  
 we applied the beam attenuation technique (BAT).

 The target thickness $\eta$ of the N$_2$ gas in the gas-cell is defined as
 
\begin{equation}\label{pi}
 \eta = \frac{\ell P}{\kappa T} \ ,
\end{equation}
 
 \noindent where $\ell$ is its effective length, $P$ the pressure and 
 $T$ the temperature of the gas cell. $\kappa$ is Boltzman's constant.
 
 The SGR method is based on the solutions to the equilibrium equations for the fraction 
 of neutral particles to the number of anions in the ion beam $F_0$ as a function of the
 target thickness $\eta$. $F_0$ was derived from
 
\begin{equation}
  F_0 = \frac{I_0 - I_b}{I_i} \ ,
\end{equation} 
 
 \noindent where $I_0$ is the signal count rate of oxygen atoms that resulted from the
 interaction with the gas, $I_b$ is the ion-beam background count rate resulting from the
 ion-beam interaction with the residual vacuum system and $I_i$ is the count rate corresponding 
 to the parent ion beam. The SGR method is described in more detail in the references 
 \cite{salazar2010,allen1995,mcdaniel1993}. In this method, the fraction $F_0$ of neutral 
 particles formed after the ion-beam loses an electron by collisions with the gas is a
 function of the initial parent ion-beam intensity and the target thickness. The 
 single-detachment cross section $^s\sigma_{-10}$ was derived from the first-order 
 approximation to $F_0$
 
 \begin{equation}\label{SGReq}
   F_0 = \ ^s\sigma_{-10}\eta ,
 \end{equation}

 \noindent 
 where $^s\sigma_{-10}$ is the single collision detachment cross section measured with 
 the SGR method and $\eta$ is the target thickness.

\begin{figure}
\begin{center}
  \includegraphics[width=\columnwidth]{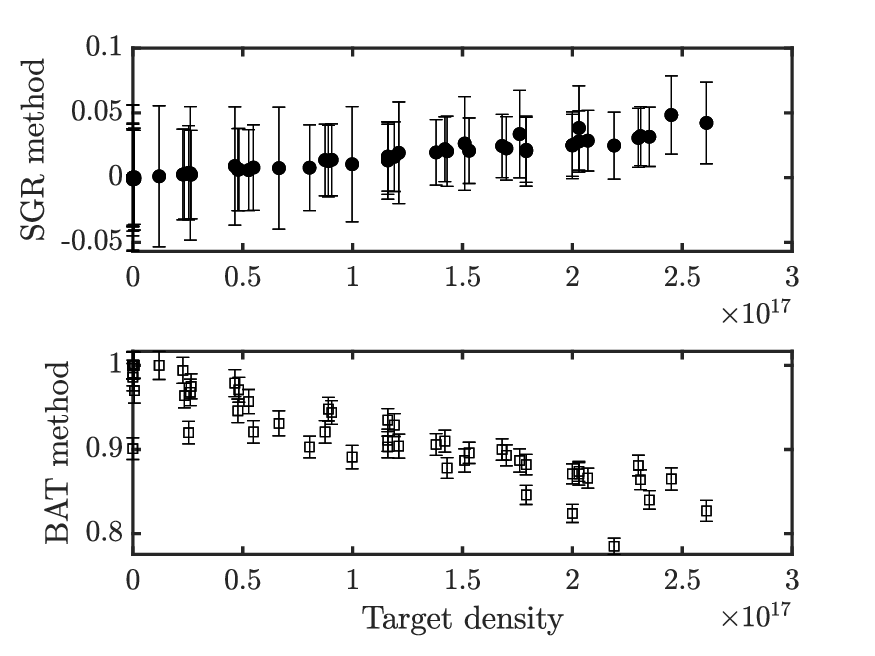}
  \caption{ \label{exampl} Examples of the measured data at 5 keV. Top panel corresponds 
  to the reduced data for the SGR method: F$_0$ as a function of the target thickness $\eta$
  in Eq. \ref{SGReq}. The bottom panel corresponds to BAT: $I/I_0$ as a function of
  $\eta$ in Eq. \ref{BATeq}.}
\end{center}  
\end{figure}

 Examples of a growth rate curve and a beam attenuation curve of the present work are shown 
 in Fig. \ref{exampl}. All measurements were carried under single collision 
 conditions, {\it i.e.}, when the functional dependence of the normalized signals 
 to $\eta$ was linear. Under these conditions, higher-order effects are expected to be negligible. 
 
 The beam attenuation technique (BAT) consists of measuring the decline of the parent 
 ion-beam intensity as a function of the gas thickness $\eta$. The cross section was
 derived from:
 
\begin{equation}\label{BATeq}
 I = I_0\exp{(- ^b\sigma_{-10}\eta)} \ ,
\end{equation} 

\noindent where $\eta$ is given by Eq. (\ref{pi}), $I$ is the ion beam intensity as a function
 of $\eta$ and, $I_0$ is the initial intensity of the ion beam. $^b\sigma_{-10}$ is the total 
 electron-detachment cross section. An example of a BAT curve for the present data is shown 
 in Fig. \ref{exampl}. The BAT technique has been used and described in more detail in the 
 references \cite{nascimento2013, ginette2014}.
 
 
 The dispersion of the data in Fig. \ref{exampl}  is mainly caused by the instability in 
 the ion-beam intensity, originated by plasma fluctuations in the ion source. In the case 
 of BAT, the vertical axes of Fig. \ref{exampl} correspond to the ion-beam intensity readings 
 normalized to $I_0$. In the case of SGR, the vertical axis correspond to the average of 
 the initial and final ion-beam currents. This could be the reason for the difference 
 in the data dispersion. Data sets where the ion-beam intensity differed by 10\% or more 
 during the beginning and the ending of an experimental measurement were discarded.
 
 Once a distribution of cross-section values was obtained at each energy, standard deviations 
 were derived as a measure of the statistical errors, that were in turn combined with the systematic 
 uncertainty to derive total errors per energy point. Our whole set of data was normalized 
 by a single constant factor so that our cross-section data point measured at 5 keV using SGM 
 coincides with the 5 keV correspondent value also measured using SGM by Matic and 
 Covic \cite{matic1971}.

                          \section{RESULTS AND DISCUSSION}

\begin{figure}
\begin{center}
 \includegraphics[scale=0.35]{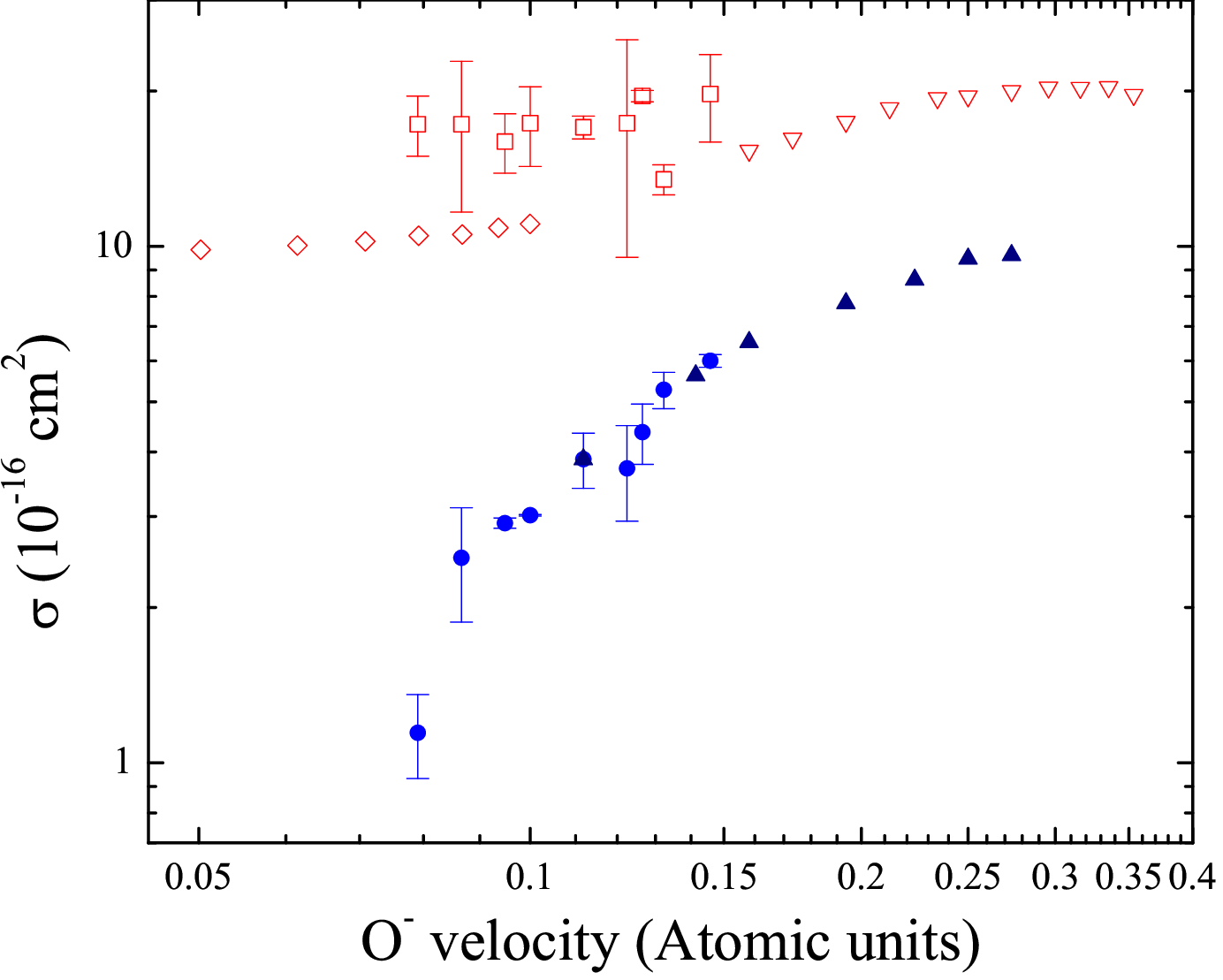}
 \caption{ \label{result1} Total electron-loss cross sections for Eq. \ref{collisEq}. 
 Data measured with BAT, open squares: present work; diamonds: Bennet {\it et al.}
 \cite{bennet1975}; inverted triangles: Tsuji {\it et al.} \cite{tsuji1989}. 
 Data measured with SGR, closed circles: present work;
 up triangles: Matic and Covic \cite{matic1971}. Error bars for the present data
 correspond to total errors, standard deviation, and systematic uncertainties. Errors
 for cited work are not plotted.}
\end{center}  
\end{figure}

 The present data are shown in Fig. \ref{result1}. Data points correspond to
 the beam attenuation technique or BAT (open data points), and to the signal growth ratio
 or SGR (closed data points). In the figure are also plotted data from Bennet {\it et al.}
 \cite{bennet1975}, measured with BAT, and data from Matic and Cobic \cite{matic1971},
 measured with SGR.

 At first approximation, BAT and SGR cross-sections are expected to have similar values, 
 given that the ion beam composition is only of O$^-$ ions and that the dependence of
 the signals with target density is under single collision conditions. However, the
 cross sections measured with each technique are very different. This kind of disagreement 
 at the low energy interval has been a question in the field and is not restricted to the 
 O$^-$ + N$_2$ collision system \cite{Rahman1986}. 
 
 It is important to note that by using BAT the total decrease of the O$^-$ ions from 
 the ion beam is measured, therefore all processes that may cause a reduction of the 
 ion-beam intensity are accounted for. By using SGR method, only neutral atoms resulting 
 from the interaction with N$_2$ are detected, which means that processes that produce 
 neutral atoms in the direction of the ion beam are accounted for.
 
 This difference may be caused by a process that is not measured in the SGR method and
 influences mainly the BAT cross-section for $^b\sigma_{-10}$ to be larger than $^s\sigma_{-10}$.  
 One aspect of the data is that the difference in the cross sections seems to decrease as 
 the interaction energy increases. We note that as the kinetic energy increases, the time 
 of flight within the apparatus becomes shorter. A possible explanation consists of the
 presence of auto-detaching states (AD) of O$^-$ that release the electron in its path to 
 the lateral CEM and in the electric field of the PPs. As the time of flight becomes longer 
 in the low energy interval of the present study, the AD states are more likely to deplete 
 the O$^-$ population in its trajectory to the detector, and as a consequence,
  $^b\sigma_{-10}$ becomes larger than $^s\sigma_{-10}$.
 
 This tendency may be consistent with the presence of a time-dependent process,
 
\begin{eqnarray}
\label{metaEq1} \mbox{O}^- + \mbox{N}_2  & \longrightarrow & ^m\mbox{O}^- + \mbox{N}_2^* \\
\label{metaEq2} ^m\mbox{O}^- & \rightsquigarrow (t) \rightsquigarrow & \mbox{O}^* + \mbox{e}^-
\end{eqnarray} 

 \noindent where the star is used to indicate an undetermined final state and the super index 
 $m$ in $^m$O$^-$ is used to designate a metastable auto-detaching state.
 Therefore, to explain the difference between the cross-sections, we propose the next hypothesis:
 Metastable states that auto-detach after the collision contribute more to $^b\sigma_{-10}$ 
 cross section because the extra electron can also be lost in its trajectory to the lateral CEM.
 
 Other processes involved in the interaction may be direct electronic detachment, electron transfer
 (ET), double detachment (DD) and single detachment followed by electron transfer where the 
 final state is O$^0$. However, the last process would imply secondary collisions and can be ruled 
 out from single collision conditions (see Fig. \ref{exampl}).

 In the case of DD channel, its contribution has been measured to be negligible for other 
 collision systems \cite{hernandez2016,ishikawa1989}, and for O$^-$ + N$_2$, in particular, 
 \cite{matic1971}. In this experiment, we measured the relative contribution of the resulting 
 O$^+$ yield by reversing the polarity of the PP showing a very low counting rate of, at most, 
 one order of magnitude lower when compared to the intensity of the SGR signal. This demonstrates 
 that the DD channel contribution cannot justify the difference. In the case of ET, this 
 process cannot be separated by these methods and its contribution is expected to have the 
 same effect in both techniques. 

 Another possibility is electric field-induced detachment. This possibility is discarded 
 on the next bases: a procedure in this experiment consisted in verifying the detectors to 
 have similar efficiencies. This was carried out by measuring the count rates with the 
 electric field on and off, and without gas load in the gas-cell. This shows that the electric 
 field alone did not induce measurable electron detachment. Consistently, the PP electric 
 field intensity resulted inchoate ($<$ 300 V cm$^{-1}$) to induce electron detachment
 \cite{halka1993}.

 Figure \ref{result1} shows that the present BAT data are consistent with data 
 from Bennet {\it et al.} \cite{bennet1975} and Tsuji {\it et al.} \cite{tsuji1989}, 
 measured with the same method. For the case of the cross sections measured 
 with the SGR method, the present data agree with the results from Matic and Covic \cite{matic1971} 
 in the coincident energy range from 5 keV to 8.5 keV. 
 The apparent disagreement between both sets of data previously published for the process of 
 Eq. (\ref{collisEq}) has been solved under the hypothesis stated above.

 We hasten to note that BAT cross-sections are consistent in the sense that their values fall within
 close order of magnitude scales. However, they disagree. Yet, this is congruent with the hypothesis 
 of auto-detaching metastable states, because the cross-section values would depend on the
 time of flight within the machine. 

 We introduce the idea that AD states $^m$O$^-$ form as a result of the collision 
 with the N$_2$ gas target. Then, these states may decay along their trajectory
 from the gas cell to the CEM detectors. When they pass through the electric field between 
 the PP analyzing plates, they continue to decay at their natural rate and may not be 
 detected by the central CEM.
  

\begin{figure}
\begin{center}
 \includegraphics[scale=0.35]{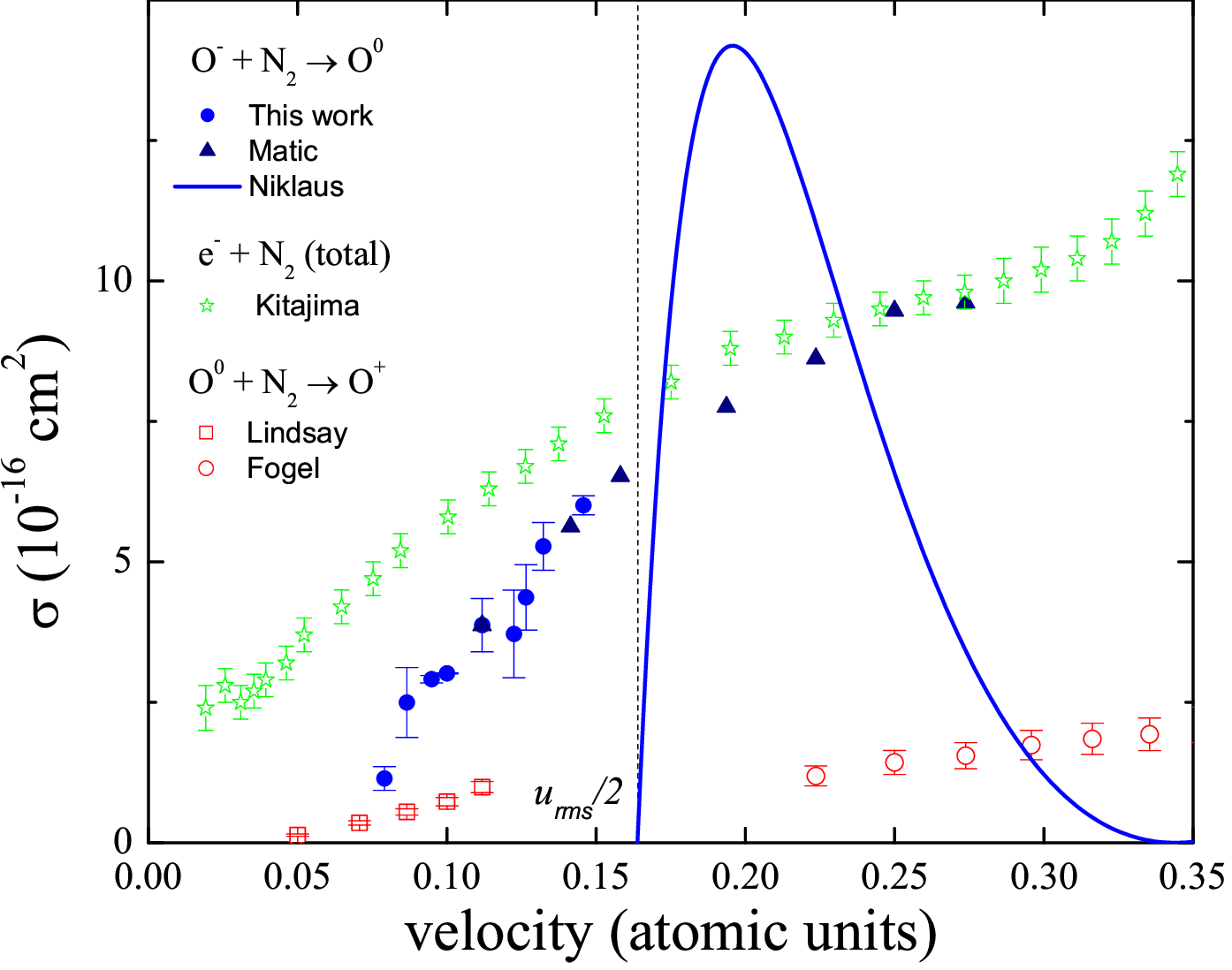}
 \caption{\label{figFCM} Cross sections for processes relevant to the free-electron-model
  hypothesis for the present interaction system.  Open stars: electron-scattering cross 
  sections e$^-$ + N$_2$ of Kitajima \cite{Kitajima2017}. Electron-detachment cross sections 
  for neutral projectiles O$^0$ + N$_2$: open squares, Lindsay \cite{Lindsay2004}; 
  open circles, Fogel  \cite{Fogel1959}. Electron-detachment cross sections for anionic 
  projectiles O$^-$ + N$_2$:  closed up triangles, Matic \cite{matic1971};  closed circles, 
  present SGR data; line, Eq.(\ref{Niklaus1}). Vertical dotted line: O$^-$ velocity given by 
  Eq. (\ref{v0Bohr}).
 }
\end{center}  
\end{figure}

  In Fig.~\ref{figFCM} we compare data for O$^-$ + N$_2$ single-electron-detachment
  cross sections measured by SGR method and with cross sections for two related 
  projectiles incident on N$_2$: electrons and atomic oxygen. The total electron 
  scattering cross sections for e$^-$ + N$_2$ agree with the data for O$^-$ + N$_2$ at 
  v $\sim$ 0.25 a.u., showing that above this velocity the description of the anion 
  as a quasi-free electron plus a neutral core is meaningful for the O$^-$ + N$_2$ 
  detachment collision channel. The comparison between O$^-$ + N$_2$ and O$^0$ + N$_2$ 
  electron-detachment cross sections also supports the quasi-free electron picture. 
  The O$^0$ + N$_2$ cross sections are much smaller than the O$^-$ + N$_2$ equivelocity ones, 
  suggesting a dominating contribution of the loosely-bound electron in an O$^-$ + N$_2$ 
  electron-plus-core simplified modeling of the anionic projectile. 

  In order to estimate projectile-electron-loss cross sections, Bohr and 
  Lindhard \cite{Bohr1948,Bohr1954} calculate the cross section for energy transfer 
  greater than T in a collision between a free electron at rest 
  (in the projectile´s frame of reference) and a heavy particle (the target) with 
  charge $Z_{ef}$, where T$_{max}$=$2mv^2$ is the upper limit for energy transfer in such 
  a collision. Thus, in this approach, the neutral target is modeled as a dressed charge 
  that ionizes the electron loosely bound to the projectile. 
  Niklaus {\it et al.} \cite{niklaus1994,Sayer1977} include in the Bohr-Lindhard 
  version of the FCM an empirical velocity-dependent effective charge $Z_{ef}(v)$ and obtain
  
\begin{equation}\label{Niklaus1}
\sigma(v)=\begin{cases}
			0, & \text{if $v < v_0$ }\\
            \sigma_0(v) \left[ 1 - \left(\frac{v_0}{v}\right)^2  \right], & \text{if $v > v_0$ }
		 \end{cases}
\end{equation}
with 
\begin{equation}\label{v0}
v_0 = \frac{u_{rms}}{2},
\end{equation} 
\begin{equation}\label{Niklaus2}
\sigma_0(v) = \pi a_B^2 Z_{ef}(v)^2   \left( \frac{1}{vv_0} \right)^2,
\end{equation} 
and
\begin{equation}\label{Niklaus3}
 Z_{ef}(v) =  Z_T \left[  1- 1.08 \exp{(-80.1~ Z_T^{-0.506} (v/c)^{0.996} )} \right]  \ ,
\end{equation} 
where $a_B$ is the Bohr radius, $c$ is the velocity of light, and  and $Z_T$ is the 
target atomic number. For anions, the Bohr-Lindhard's threshold velocity is parameterized
in terms of the Electron Affinity (EA) \cite{Kristiansson2022,ning2022} as
\begin{equation}\label{v0Bohr}
v_0 =\frac{1}{2}\sqrt{\frac{EA(eV)}{13.6}}.
\end{equation}

In Fig.~\ref{figFCM} we also show the results from Eq.~(\ref{Niklaus1}) for O$^-$ + N$_2$.
The agreement with experimental data is poor, for two reasons: (i) The modeling of 
the target as a dressed charged particle is a naive approximation to a much more complex
problem \cite{montenegro1994}, and is especially inadequate for collision 
between an electron and the N$_2$ target \cite{song2023}. 
(ii) The value for the free-collision threshold 
is overestimated due to neglecting the velocity of the 
ionized electron within the projectile frame of reference. Variations of the FCM deal 
with these limitations by including a combination of previously known cross sections for 
electron-impact elastic scattering by the target and the projectile 
electronic-velocity distribution (e.g. \cite{bates1966,bates1967,bates1969,risselmann1992}). 

This more general FCM approach includes a velocity threshold for the electron detachment, 
since there is still a maximum momentum that the massive target can transfer to the 
projectile's electron. This limit is now affected by the angular dependence of elastic 
scattering of a free electron by the target and by the electronic distribution of the 
anionic projectile \cite{risselmann1992,sigaud2008,sigaud2011}. An onset 
for detachment cross sections is expected as the projectile velocity increases, 
although a velocity threshold convoluted with the projectile internal velocity distribution. 

However, quadruple integrals involved if this approach are a limitation to FCM 
calculations of cross sections involving anionic projectiles and molecular targets, 
often restricted to the intermediate-to-high velocity regime ({\it e.g}~\cite{nascimento2013,ginette2014,risselmann1992,sigaud2008,sigaud2011,Heinemeier1976,
santanna2004,santanna2009}). In the present work, we combine the 
Bohr-Lindhard's approach with a threshold analysis of the general Risselman's approach 
in order to obtain a simple analytical expression of the FCM threshold behavior 
that takes into account the electronic velocity distribution within the projectile.



In Risselmann's formulation of the FCM \cite{risselmann1992}, the total-electron-loss cross 
section $Q(v_N)$ for a one-electron projectile for which the nucleus has a velocity 
$\mathbf{v}_N = v_N \hat{z}$ is given by
\begin{equation}\label{sigmaFCM2}
 \int_0^\infty f(u)du \int_0^\pi \frac{1}{2} \sin(\beta) d\beta  \int_0^{2\pi} d\phi \int \sigma(v,\theta)  \sin(\theta)d\theta  \ ,
\end{equation} 

\noindent where $u$ and $\beta$ are, respectively, the modulus and the angle with $z$ axis of the 
projectile electron in the projectile frame of reference, and $v$ is the velocity of the 
same electron, but in the laboratory frame of reference. $f(u)$ is the distribution of 
absolute values of the velocity of the projectile's electron. 
$\sigma(v,\theta)$ is the differential 
electron-scattering cross section at an angle $\theta$ for a free electron impinging on the target. 
For elastic collisions, $\Delta E=0$, Risselmann's expression of the inequality regarding energy and momentum 
conservation reduces to

\begin{multline}\label{eqthreshold}
(v_N+u\cos(\beta))(1-\cos(\theta)) + u\sin(\beta)\cos(\phi)\sin(\theta) \\ \geq  u_{rms}^2/(2v_N),
\end{multline}
 
\noindent where the characteristic anion velocity $u_{rms}$ is obtained from EA$=mu_{rms}^2/2$. 
While $u_{rms}$ is often interpreted as a root mean squared value, it is normally obtained 
from experimental EA values \cite{risselmann1992,sigaud2008,sigaud2011}. Near threshold, close 
collisions are most relevant to higher momentum transfer. Thus, we approximate $\theta \approx \pi $ 
and, using $v_0=u_{rms}/2$, obtain from Eq. (\ref{eqthreshold})

\begin{equation}\label{eqSimplethreshold}
\frac{v_N}{v_0} \left( \frac{v_N}{v_0}+\frac{u}{v_0}\cos(\beta) \right) \geq  1 .
\end{equation}

If the electron velocity in the projectile frame of reference ($u$) is negligible compared to 
the anion nucleus velocity, then  $v_N \geq  v_0 $ and $v_0$ is the threshold 
velocity (as in the Bohr-Lindhard's approach). 
Otherwise, Eq. (\ref{eqSimplethreshold}) 
results in a threshold velocity that is a function of the product $u \cos(\beta)$, 
which is the $z$ component of the electron velocity within the anion, $u_z$. Defining 
$s=v_N/v_0$ and $t=u/v_0$ We can write


\begin{equation}\label{eqinequalityst}
t \cos(\beta)\geq  \left(\frac{1}{s} - s \right)
\end{equation}

The particular case $\cos(\beta)=\pm 1$ gives electron momentum 
parallel or anti-parallel to the anion's nucleus velocity, resulting in the $s$ positive 
limit solutions 
\begin{eqnarray}
\label{degree2a} s_< = \sqrt{  1+ \left(\frac{t}{2} \right)^2   } -\frac{t}{2},  \\
\label{degree2b} s_> = \sqrt{  1+ \left(\frac{t}{2} \right)^2   } + \frac{t}{2}.
\end{eqnarray} 
If the distribution $f(u)=\delta(u-u_{rms})$ then $t=2$, and therefore $s_< = \sqrt{2}-1 $ 
and $s_> = \sqrt{2}+1 $. Thus, within the $\delta$ approximation for $f(u)$,
\begin{equation}\label{eqvlimit}
v_< =\left( \sqrt{2}-1 \right) v_0~ \approx~ 0.414~\frac{1}{2}\sqrt{\frac{EA(eV)}{13.6}}.
\end{equation}

The ratio R between cross section for projectile electron loss $\sigma_{O^-}$ and total electron 
elastic cross sections $\sigma_e$ can be used to compare experimental data to the FCM results. Within the 
Bohr-Lindhard's approach, assuming $\sigma_0=\sigma_e$, this leads to an expression 
unaffected by the dressed-charge approximation to the target:
\begin{equation}
\label{eqRBL}
 R_{B-L} =  1 - \left(\frac{v_{0}}{v}\right)^2.
\end{equation}
We substitute the Bohr-Lindhard's threshold velocity by $v_<$ (Eq. \ref{eqvlimit}) and obtain
\begin{equation}
\label{eqR}
 R =  1 - \left(   \frac{  \left(\sqrt{2}-1\right)   ~v_{0}}{v}   \right)^2.
\end{equation}

\begin{figure}
\begin{center}
 \includegraphics[scale=0.35]{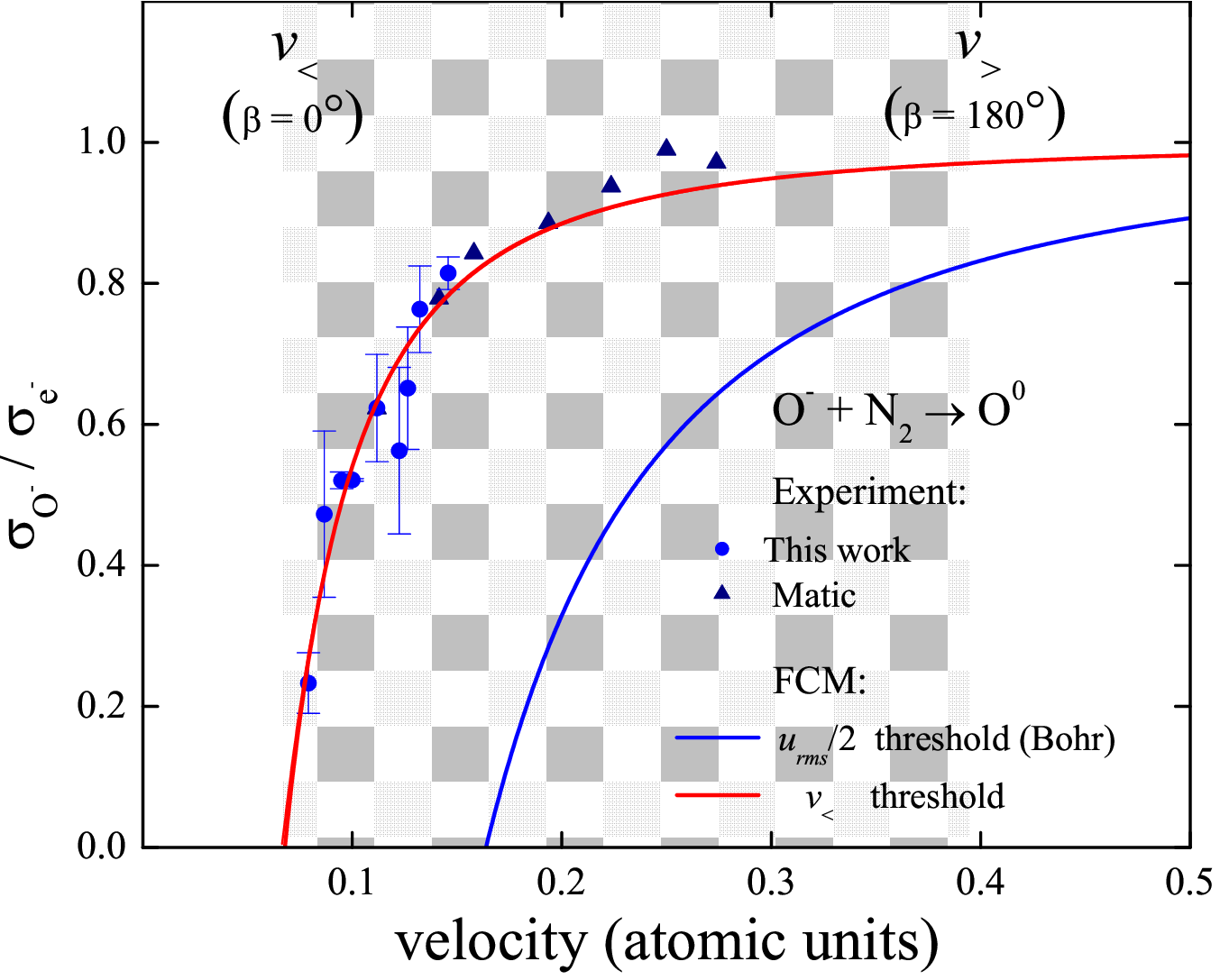}
 \caption{ \label{figRatio} Ratio of cross sections for O$^-$ + N$_2$ projectile electron 
 loss and total electron scattering by N$_2$ (using e$^-$+N$_2$ of Kitajima \cite{Kitajima2017}). 
 Closed circles, present SGR data; closed up triangles, Matic and Cobic \cite{matic1971}; 
 Blue line, Bohr-Lindhard FCM; Red line, modification of Bohr-Lindhard FCM 
 with the $v_<$ velocity threshold; The grey shaded area highlights the velocity range with threshold 
 affected (within the FCM) by the angle $\beta$ between the incidence direction and projectile 
 electron direction. The region is delimited by $v_<$ ($\beta=0$) and $v_>$ ($\beta=180\deg$).}
\end{center}  
\end{figure}

Figure \ref{figRatio} shows ratios of O$^-$+$N_2$ projectile electron loss to the e$^-$ + N$_2$ total electron 
elastic cross sections of Kitajima {\it{et} \it{al}.} \cite{Kitajima2017}. For sufficiently 
high velocities, the FCM predicts a ratio of 1. The ratios for the higher-energy data of 
Matic \cite{matic1971}, with velocities in the range between $v_0$ and $u_{rms}$, are close 
to one. 

We show in Fig.~\ref{figRatio} the FCM results obtained from the B-L approach (Eq. \ref{eqRBL}) and 
and the B-L result with threshold velocity $v_0$ replaced by the $v_<$ expression obtained in 
the present work (Eq. \ref{eqR}). We represent the range of velocities with $\beta$ integration 
restrictions (Eq. \ref{eqinequalityst}) by the grey shaded area. The comparison between the two FCM 
results and the experimental data, both from this work and from Matic and Cobic \cite{matic1971}, 
shows that the substitution of the threshold velocity $v_0$ by $v_<$ results in good FCM 
description of the anionic-projectile collision system studied in the present work. 
It also suggests that for anionic projectiles (for which the EA values are at most of a few eV), 
even close to the FCM threshold, the $\beta$ distribution are more relevant to cross-section convolution
that the distribution of absolute values $u$ (see Eq. \ref{sigmaFCM2}). 
The parametrization of R only on EA, 
via $v_<$, provides a contribution in the pursue of scaling laws common to 
few-keV projetile-electron-loss cross sections for anionic 
projetiles ({\it e.g.} \cite{zappa2004,wu2007,geng2013,min2017,yang2022}).

%
%
%
%


It is important to note that the existence of the threshold for quasi-free electron 
detachment doesn't imply that 
the projectile-electron-detachment cross section is expected to be zero below 
threshold. At very low velocities, molecular states formed by projectile and target 
can result in large cross sections \cite{esaulov1986}. In the case of O$^-$+N$_2$ 
those states are related to the temporary formation of a N$_2$O$^-$ 
complex \cite{comer1974}. The velocity dependence of projectile-electron-detachment 
cross sections with a peak before threshold was observed, for example, 
in O$_2^-$+N$_2$ collisions \cite{mendes2019}.

%
%
%
%
%

\section{CONCLUSIONS} 

 The total projectile-electron-loss cross sections for O$^-$+N$_2$ in the energy range 
 from 2.5 to 8.5 keV have been measured with two methods: the beam attenuation 
 ($^b\sigma_{-10}$) and the signal growth rate techniques ($^s\sigma_{-10}$). 
 Close to this energy range, previously published cross sections show a long-standing 
 considerable disagreement, depending on the measurement method used.
 The present data have also shown this discrepancy between results originated in different 
 methods. At the same time, for each method, the present results are consistent with the 
 cross sections measured with corresponding techniques. We propose that this discrepancy can be 
 explained by the presence of auto-detaching metastable states that cause $^b\sigma_{-10}$ 
 to be larger. This hypothesis may solve the aforementioned disagreement.
 
 From a fundamental point of view, a negative ion can be considered a simple carrier to a free 
 electron under certain circumstances. The free collision model (FCM) is based on this assumption.
 However, the validity of this model close to the velocity threshold  
 has been rarely assessed. In this study, we present a general, simple, and analytical expression 
 for the velocity threshold within the FCM, easily applied for anionic projectiles. 
 When applied to the O$^-$+N$_2$ collision system, the expression shows consistency 
 with the experimental threshold value obtained from the velocity dependence of our data.
 
 In addition to its fundamental interest, in plasma modeling, electron scattering is a very 
 important process given the abundance
 of electrons and the impact of this process on the plasma electron density function. In this work,
 we demonstrate that the process of electron loss from O$^-$ at a speed $v$ above $\sim$ 
 0.25 a.u. the cross section for electron detachment is as relevant as electron scattering
 and in a given plasma environment with a high density of negative ions, such as carbohydrate plasma,
 electron detachment could be even more relevant than electron scattering. 

 \section*{Acknowledgements}
 
 G. H. acknowledges the National Autonomous University of Mexico (UNAM) through the
 Support Program for the Improvement of the Academic Staff of UNAM (PASPA). Technical 
 support from Guillermo Bustos, H\'ector H. Hinojosa, Reyes Garc\'ia, Armando Bustos, 
 Juana A. Romero and Arturo Quintero is acknowledged. This study was supported in part 
 by the Brazilian agency Funda\c{c}\~{a}o de Amparo \`{a} Pesquisa do Estado 
 do Rio de Janeiro (FAPERJ), code E-262109342019 and in part by CONAHCYT CF-2023-I-918.
 
\bibliography{References}

\end{document}